\documentclass[
 reprint,
 amsmath,amssymb,
 aps,
]{revtex4-2}

\usepackage[T1]{fontenc}
\usepackage{graphicx}
\usepackage{dcolumn}
\usepackage{bm}
\usepackage{booktabs}
\usepackage{float}
\usepackage{dsfont}
\usepackage{calrsfs}
\usepackage{xcolor}
\usepackage{pict2e}
\usepackage[percent]{overpic}
\usepackage{upgreek}
\usepackage{xfrac}
\usepackage{abraces}
\usepackage{braket}
\usepackage{stmaryrd}
\usepackage{empheq}
\usepackage{multirow}
\usepackage{comment}
\usepackage{listings}

\DeclareMathAlphabet{\pazocal}{OMS}{zplm}{m}{n}
\usepackage[colorlinks=true,linkcolor=blue,citecolor=blue,urlcolor=blue]{hyperref}

\begin{document} 

\preprint{APS/123-QED}

\title{Constraints on Galactic Dark Photons from a DALI Prototype}
\author{Javier De Miguel$^{1,2,3}$}
 \email{jdemiguel@iac.es}


\author{Antonios Gardikiotis$^{4,5}$}%
\author{Elvio Hern\'andez-Su\'arez$^{1}$}%
\author{Roger J. Hoyland$^{1}$}%
\author{Enrique Joven$^{1}$}%
\author{Abaz Kryemadhi$^{6}$}%
\author{Haroldo Lorenzo-Hern\'andez$^{1}$}%
\author{Marios Maroudas$^{7}$}%
\author{Chiko Otani$^{3}$}%
\author{J. Alberto Rubi\~no-Mart\'in$^{1,2}$}%
\author{Yannis K. Semertzidis$^{8}$}%
\author{Michael E. Tobar$^{9}$}%
\author{Konstantin Zioutas$^{4}$}%

\collaboration{The DALI Collaboration}

\affiliation{$^{1}$Instituto de Astrof\'isica de Canarias, E-38200 La Laguna, Tenerife, Spain}

\affiliation{$^{2}$Departamento de Astrof\'isica, Universidad de La Laguna, E-38206 La Laguna, Tenerife, Spain}

\affiliation{$^{3}$RIKEN Center for Advanced Photonics,
519-1399 Aramaki-Aoba, Aoba-ku, Sendai, Miyagi 980-0845, Japan}

\affiliation{$^{4}$Physics Department, University of Patras, GR 26504, Patras-Rio, Greece}

\affiliation{$^{5}$Institute of Quantum Computing and Quantum Technology, National Centre for Scientific Research ``Demokritos'', 153 41 Athens, Greece}

\affiliation{$^{6}$Computing, Math \& Physics, Messiah University, Mechanicsburg, PA 17055, USA}

\affiliation{$^{7}$Institute for Experimental Physics, University of Hamburg, 22761 Hamburg, Germany}

\affiliation{$^{8}$Innovative Solutions R\&D LLC, Stony Brook, NY 11790, USA}

\affiliation{$^{9}$Quantum Technologies and Dark Matter Research Lab, Department of Physics, University of Western Australia, Crawley, WA 6009, Australia.}

\date{\today}

\begin{abstract}
An analysis of 36 hours of data from a DALI haloscope reveals no statistically significant excess attributable to dark-photon dark matter. We therefore set new constraints in the 6.88691--6.91792 GHz band, reaching a dark-photon-to-photon kinetic mixing strength of $\chi\lesssim 6.8\times10^{-14}$ at $28.54 \, \upmu\mathrm{eV}$. To our knowledge, this result established the strongest laboratory-based exclusion limit in this frequency range.

\end{abstract}

\maketitle



\noindent\textbf{Introduction.} 
Dark photon, also referred to as hidden photon or paraphoton, is a hypothetical gauge boson that mixes kinetically with standard photons \cite{Okun:1982xi}. The interaction term relevant for this work is
\begin{equation}
\pazocal{L}^{\mathrm{int}}_{\gamma'\gamma}= 
-{\frac {1}{2}}  F_{\mu {\nu} } X^{\mu \nu } \chi \,,
\label{Eq.9}
\end{equation}
where we denote by $X^{\mu \nu}$ the field strength tensor of the dark photon field;  $\chi$ being the dimensionless kinetic mixing strength.  

Dark photon has been hypothesized to constitute dark matter, an elusive substance whose existence is supported by indirect evidence \cite{1970ApJ...159..379R}. A promising experimental approach to probing cosmic dark matter is the haloscope, first proposed by Sikivie \cite{1983PhRvL..51.1415S}, which in its most common implementation employs a resonant cavity to enhance the weak microwave signal induced by virialized axions and dark photons in the Galactic halo. In this manuscript we report the first dark photon search from a proof-of-principle prototype of the Dark-photons \& Axion-Like particles Interferometer (DALI), a new-generation DM haloscope designed to complement axion, dark photon, and gravitational wave searches at high frequencies~\cite{DeMiguel2021,DeMiguel:2023nmz,Cabrera2023qkt,2024JInst19P1022H,PhysRevD.110.072013,DeMiguel:2024cwb, DeMiguel:2026mvi}.
\begin{figure}[b]
    \centering    \includegraphics[width=.4\textwidth,trim=0cm 0 0cm 0,clip]{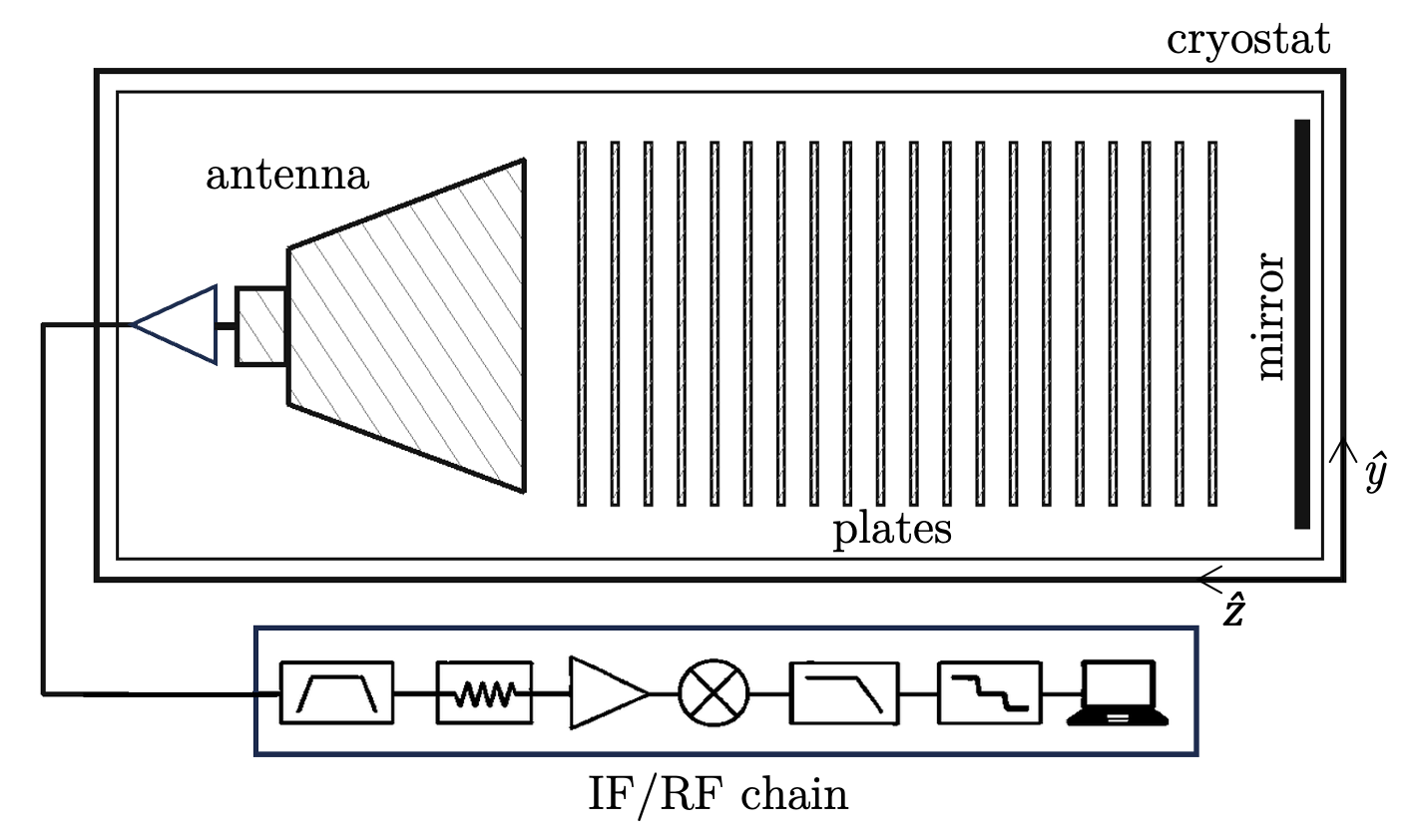}
    \caption{Schematic view of the DALI(PoP) prototype. Ambient dark photons are converted into microwave photons, which are collected by the antenna. The resonator, composed of twenty ZrO$_2$ plates and enclosed by a copper mirror, enhances the weak paraphoton-induced signal to an observable level. The detected signal is subsequently processed by the readout chain shown below (from left to right: cryogenic amplifier, band-pass filter, attenuator, amplifier, downconverter, low-pass filter, analog-to-digital converter, and data storage). The haloscope is housed in a Faraday cage to suppress external spurious signals (not shown). The reference axes $\hat{y}$ and $\hat{z}$ are shown, while $\hat{x}$ is the remaining direction.
}
    \label{fig_0}
\end{figure}

DALI searches for wavelike DM with a tunable Fabry--P\'erot (FP)~\cite{1899ApJ.....9...87P}
interferometer: a stack of ceramic plates whose
resonant field is sampled in the radiative near field by an antenna array,
which collects the signal over the full open-cavity aperture. Unlike a closed
cavity, the resonance frequency is set by the plate spacing and is decoupled
from the transverse area, so the resonator cross section can be scaled
independently of frequency. Dark photons convert within the electromagnetic discontinuity
without an external magnetic field, while the same apparatus probes axions
when magnetized. This architecture therefore offers a complementary route to
resonant searches at high frequencies, where conventional cavity haloscopes
struggle to retain large effective volumes. DALI relies on mature,
commercially available components, enabling rapid and cost-effective
development.

Here we report a proof-of-principle (PoP) test at reduced scale, designed to
retain part of DALI's physics reach while relieving hardware demands. The rest of this Letter is structured as follows. The section \textit{Experimental setup} describes the DALI(PoP) prototype employed to search for dark-photon DM, while \textit{Data analysis} details the observations and their statistical treatment. A new limit on the dark photon parameter space is presented in \textit{Results}.

\noindent\textbf{Experimental setup.} The DALI(PoP) prototype (Fig.~\ref{fig_0}) is hosted at the IAC
($28^\circ28'29''$~N, $16^\circ18'37''$~W) and can operate at 6--8~GHz and
$\sim$30--40~GHz. It scales DALI down by halving the number of FP layers and
reducing the plate side to $\sim$1:10, a factor $\sim$100 in area. The
full-scale instrument covers its larger aperture with a phased array of
antennas combined coherently; at the reduced aperture of the prototype a
single antenna suffices, preserving the same operating principle.
\begin{widetext}
\begin{equation}
\begin{split}
\chi \gtrsim{}&
4\times10^{-14}
\left(\frac{\mathrm{SNR}}{3}\right)^{1/2}
\left(\frac{10^{3}}{Q_L}\right)^{1/2}
\left(\frac{1+\beta}{\beta}\right)^{1/2}
\left(\frac{100\,\mathrm{cm}^{2}}{A_{\eta}}\right)^{1/2}
\\[1ex]
&\times
\left(\frac{m_{\gamma'}}{50\,\upmu\mathrm{eV}}\right)^{1/4}
\left(\frac{1\,\mathrm{day}}{t}\right)^{1/4}
\left(\frac{T_{\mathrm{sys}}}{50\,\mathrm{K}}\right)^{1/2}
\left(\frac{0.45\,\mathrm{GeV\,cm}^{-3}}{\rho_{\gamma'}}\right)^{1/2}
\left(\frac{1/3}{\kappa^{2}}\right)^{1/2}\,.
\end{split}
\label{Eq.1}
\end{equation}
\end{widetext}
Equation~\ref{Eq.1} gives the projected sensitivity, obtained by combining the
resonant signal power with the Dicke radiometer
relation~\cite{1946RScI...17..268D},
${\rm SNR}\propto P_{\gamma'}T_{\rm sys}^{-1}(t/\Delta\nu_{\gamma'})^{1/2}$.
Here $Q_L$ is the loaded quality factor, $\beta$ the receiver coupling,
$A_{\eta}$ the projection of the electromagnetic mode onto the dark-photon source
distribution, $T_{\rm sys}$ the system noise temperature, $t$ the integration
time, $\Delta\nu_{\gamma'}$ the signal linewidth, and $m_{\gamma'}$ the
dark-photon mass. The local DM density is set to
$\rho_{\gamma'}=0.45^{+0.03}_{-0.09}$~GeV~cm$^{-3}$~\cite{Staudt:2024tdq},
assuming dark photons saturate it. Throughout, the dark-photon polarization is
taken to be randomly oriented and uncorrelated over an observation long
compared with the coherence time, so that the overlap with the antenna
response averages to $\kappa^2\equiv\langle\cos^2\theta\rangle\simeq1/3$ \cite{Caputo:2021eaa}.

\begin{figure}[b]
    \centering    \includegraphics[width=.39\textwidth]{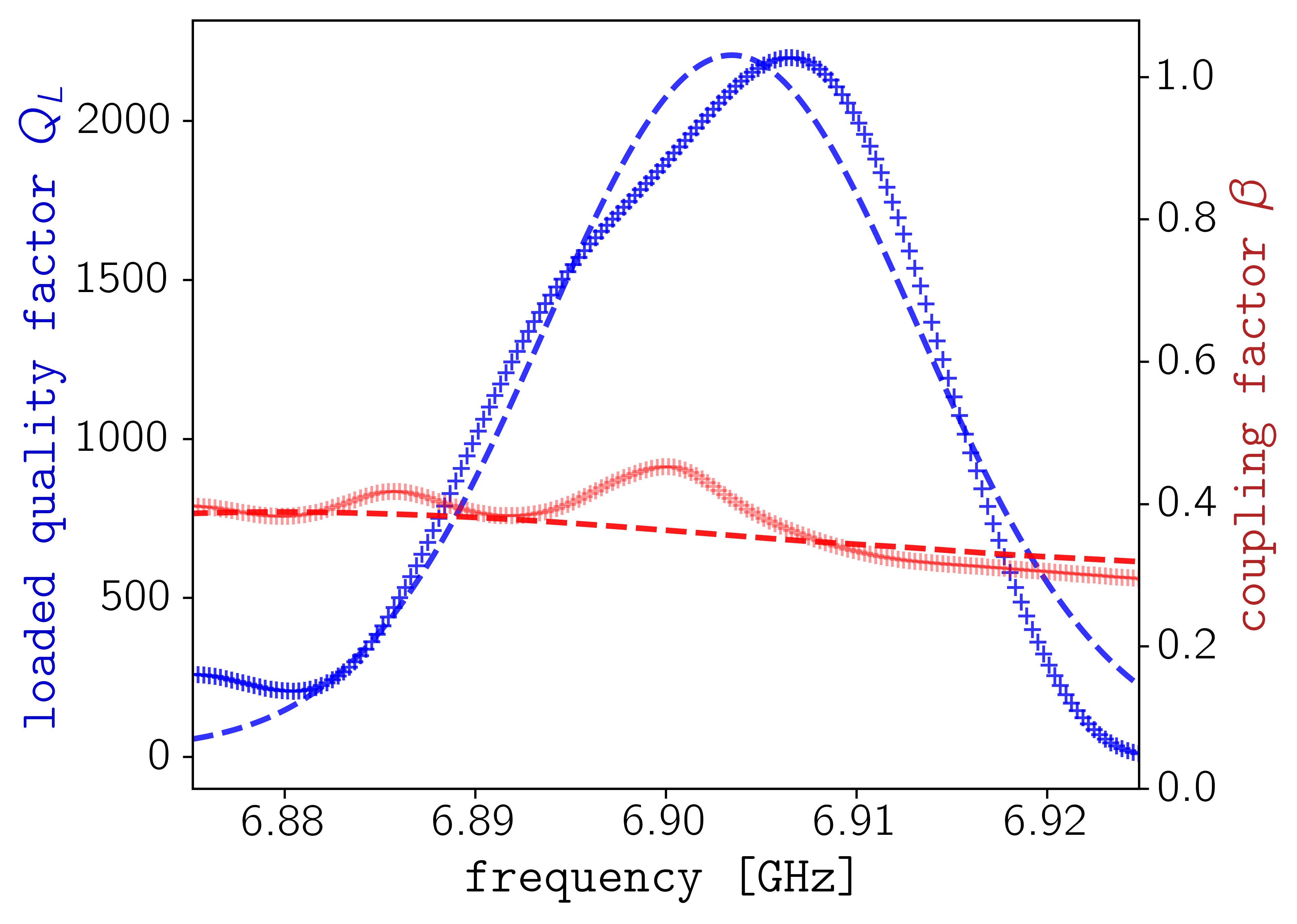}
    \caption{Loaded quality factor and antenna coupling. Blue: measured $Q_L$; the blue
dashed curve is a fit, of full width at half maximum $\sim$20~MHz. Red:
measured coupling coefficient $\beta$, the red dashed curve is a Savitzky--Golay
smoothing ($W=201$, $d=1$) of the broadband $\sim$5--8~GHz sweep, shown for
reference only. With $\beta\simeq0.36$ at resonance peak the receiver is
undercoupled. Throughout this work $Q_L$ and $\beta$ are taken from the
measured data, not from the fitted or smoothed curves.}
    \label{fig_4}
\end{figure}

The FP resonator comprises 20 yttria-stabilized zirconia (ZrO$_2$) plates,
$100\times100$~mm ($\pm0.5$~mm), 1~mm thick ($\pm0.03$~mm), surface roughness
below 0.1~$\upmu$m, $\varepsilon_r\sim30$ and $\tan\delta\sim10^{-4}$ at
centimeter wavelengths. In this pilot run the plate spacing is fixed at
$6.21\pm0.05$~mm, a single frequency step. Following
Ref.~\cite{PhysRevD.110.072013}, $Q_L$ was measured with a Rohde \& Schwarz
ZNB~20 vector network analyzer (VNA). At the present spacing
($\sim\lambda/8$, with $\lambda$ the wavelength) the response peaks at
$Q_L\approx2200$ near 6.907~GHz and follows a pseudo-Lorentzian profile
(Fig.~\ref{fig_4}). In the $\lambda/8$ mode the stack serves as the tuning mechanism, while the mirror acts as the emitting source.
Microwave absorbers (EA-PF3000-24 and 3650-40-ML) buffer spillover and
suppress reflections. A WR-137 horn antenna feeds a HEMT radiometer: a
cryogenic LNF-LNC4-8C low-noise amplifier ($\sim$40~dB gain, noise temperature
of a few kelvin) followed by a room-temperature backend with a second
LNF-LNC4-8C, a 12~dB attenuator
(Mini-Circuits\textsuperscript{\textregistered} 15542) to avoid compression at
the second gain stage, and 6--8~GHz band-pass filtering. Cooling is provided
by two closed-cycle $^{4}$He cryocoolers. The system temperature was
determined by the standard Y-factor method, heating a microwave absorber with
a resistor (Ohmite HS100 330R~J) over $\Delta T\sim60$~K, giving
$T_{\rm sys}\simeq59$~K (49.3~K from the antenna, cooling-limited under the $>$300~kg hardware load, and 9.7~K of receiver noise).

\begin{figure}[h]
\includegraphics[width=0.8\columnwidth]{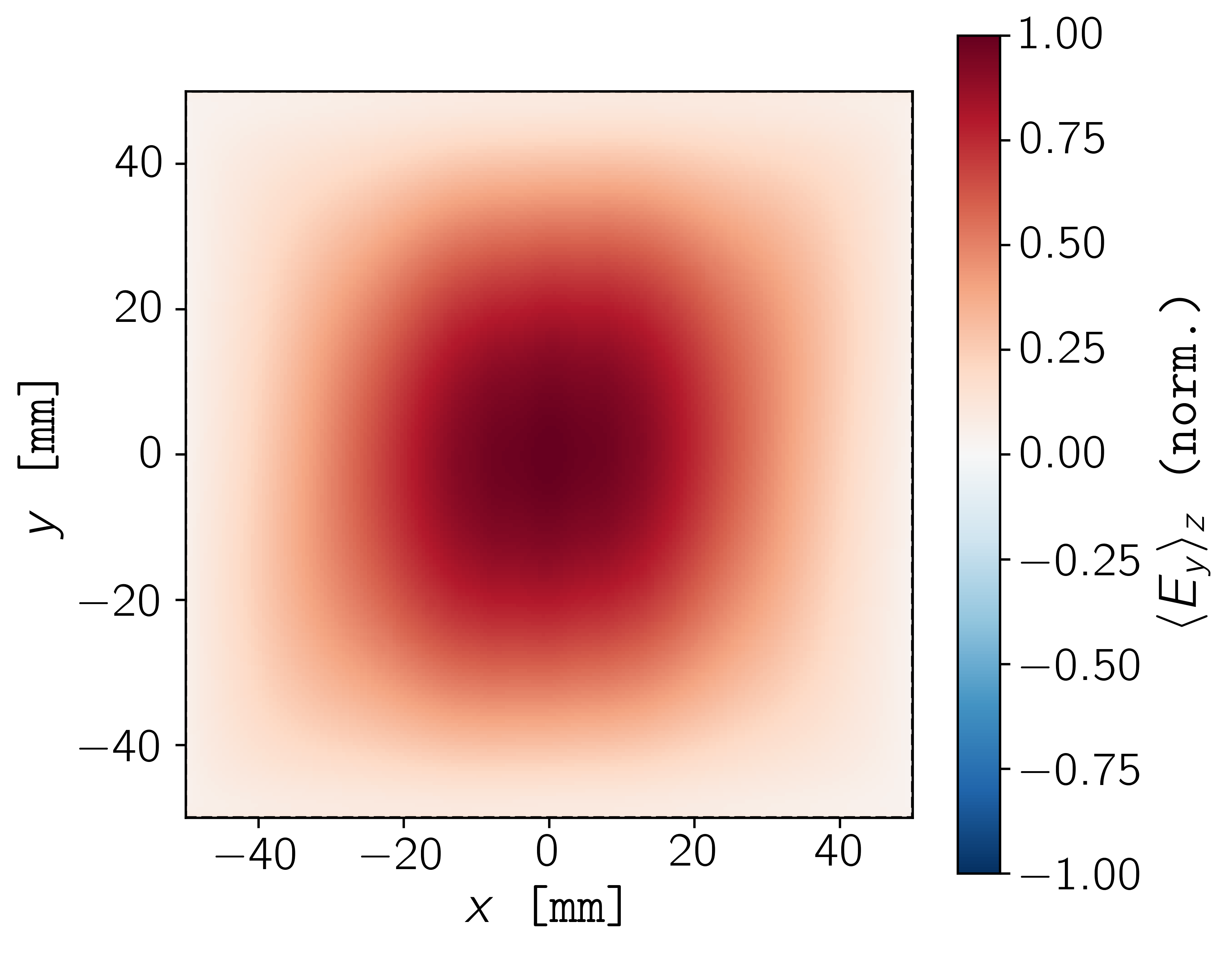}
\caption{Transverse profile of the resonant eigenmode of the $\lambda/8$
stack, averaged over the longitudinal planes of the assembly and
normalised to its maximum.  The contour marks the $100\times100$~mm$^{2}$
plate.  The coherent projection is $\eta=\bigl|\int_A\mathbf{E}_{y}\,dA\bigr|^{2}
\big/\bigl(A_{\mathrm{phys}}\int_A|\mathbf{E}_{y}|^{2}\,dA\bigr)\simeq0.62$ over those planes, with an interquartile range
$0.58$--$0.66$.}
\label{fig_3}
\end{figure}

The effective cross section is obtained from the simulated (CST~v2024) mode projection, 
$A_{\eta}\approx0.61\,A_{\rm phys}$ (Fig.~\ref{fig_3}), averaged along the stack length. The coupling is extracted from free-space VNA
measurements; $\beta<1$ (Fig.~\ref{fig_4}) indicates an undercoupled
antenna--resonator arrangement, leaving room for improved matching in future runs. Parameters
and uncertainties are listed in Table~\ref{table9}.

\begin{table}[b]
\caption{Uncertainties are weighted according to their order in Eq.~\ref{Eq.1} to estimate the propagated cumulative uncertainty in this run.}
\label{table9}
\begin{tabular}{ccc}
\\ \toprule
Parameter & Value & Uncertainty \\\midrule
Loaded quality factor ($Q_L$) [peak] & 2200 & 8\%\\
Coupling ($\beta$) [at resonance] & 0.36 & 1\%\\
Effective area ($A_{\eta}$) & 62 cm$^2$ & 6\%\\
System temperature ($T_{\mathrm{sys}})$ & 59.0 K & 2\%\\\hline
Propagation to $\chi$ sens. & & 5\%\\
\bottomrule
\end{tabular}
\end{table}

The cryostat sits inside a Faraday cage (Holland Shielding Systems B.V.) whose
connector plate carries radio-frequency filters, waveguides, and ventilation
honeycombs, attenuating external signals by more than 60~dB. Outside the cage,
a zero intermediate frequency downconverter (thinkRF\textsuperscript{TM}
R5550-408) shifts the radiometer output to baseband as in-phase and quadrature
(IQ) channels, allowing instantaneous acquisition of the full band except close
to zero hertz. The IQ outputs are low-pass filtered (SLP-36+, $-3$~dB at
$\sim$40~MHz) and digitized simultaneously by a dual-channel 14-bit
analog-to-digital converter (GaGe CSE8329); a mask balances the two channels in
amplitude and phase before combination, calibrated with a swept VNA measurement
at the downconverter input. Samples are streamed over a PCIe link to a local
server with 18~TB of storage for post-processing.

\noindent\textbf{Data analysis.} 
The datasets used to search for cosmic axions \cite{DeMiguel:2026mvi} are reinterpreted here. Data were acquired over 36 hours of effective integration time between
February 17 and March 5, 2026, using the haloscope described in the
previous section. A total of $N=12960$ spectra were recorded at
125~MS/s, centered at 6.90~GHz. Each stored spectrum is the average of
9540 consecutive dual-channel fast Fourier transforms of length
$2^{17}$ samples, giving an effective integration time of 10~s per file
and a frequency resolution $\Delta\nu_b \simeq 954$~Hz. This amounts to
approximately 64.8~TB of raw time-domain data, compressed to 44.2~GB in
the stored frequency-domain dataset.

We follow the methodology and nomenclature of Ref.~\cite{Cabrera2023qkt},
where the analysis of Ref.~\cite{PhysRevD.96.123008} was adapted to a
DALI haloscope and validated through end-to-end Monte Carlo simulations:
a synthetic dark-photon signal was injected into a noisy background and
its recovery rate used to fine-tune the prefactor of
Eq.~\ref{Eq.1}. Time-to-frequency conversions were computed with the
\texttt{Fastest Fourier Transform in the West} library~\cite{1386650}.


Spectra were cropped to the region with $Q_L\gtrsim600$, retaining the
sensitive resonant band, and averaged in the frequency domain. The
averaged spectrum was smoothed with a Savitzky--Golay (SG) baseline of
window $W=251$ and polynomial degree $d=4$. Spurious radio-frequency
interferences were then identified iteratively in each normalized
spectrum as bins exceeding $5.5\,\sigma$ relative to that spectrum's own
mean and standard deviation. Compromised bins were replaced by random
draws from the corresponding Gaussian distribution until no further
outliers appeared~\cite{doi:10.1126/sciadv.abq3765}, and their
frequencies were masked in the subsequent analysis. Residual broad
spectral structure was removed from each individual spectrum with a
second SG filter of identical parameters.


Processed spectra were rescaled bin by bin by the effective quality
factor $Q_{\rm eff}=Q_L\beta/(1+\beta)$, which accounts for the
fraction of the stored power delivered to the receiver through the
antenna. The rescaled spectra were combined and rebinned with
$K_r=3$, $K_g=5$ and $z=0.65$, chosen so that the expected dark-photon
line spans approximately $K_g$ bins. The resulting grand spectrum (GS),
of bin width $\Delta\nu_{\rm GS}\simeq2.86$~kHz, is a weighted sum
accounting for the laboratory-frame dark-photon line shape,
\begin{multline}
f(v) = \frac{2}{\sqrt{\pi}}
\left( \sqrt{\frac{3}{2}} \frac{1}{r\, v_{\gamma'}
\langle \upbeta_{\rm rms}^2 \rangle} \right)
\sinh\left( \frac{3r}{\sqrt{2}}
\frac{(v - v_{\gamma'})}{v_{\gamma'}
\langle \upbeta_{\rm rms}^2 \rangle} \right) \\
 \exp\left[ -\frac{3}{2}
\left( \frac{(v - v_{\gamma'})}{v_{\gamma'}
\langle \upbeta_{\rm rms}^2 \rangle} \right)^2
- \frac{3r^2}{2} \right] \,,
\end{multline}
where $v$ is the dark-matter velocity in the laboratory frame,
$v_{\gamma'}$ the mean dark-photon velocity in the Galactic rest frame,
$r \equiv v_{\mathrm{lab}}/v_\mathrm{rms}$ encodes the laboratory motion
relative to the halo, and $\langle \upbeta_{\rm rms}^2 \rangle =
\tfrac{3}{2}\,v_\mathrm{rms}^2/c^2$ with $v_\mathrm{rms} \simeq
270\,\mathrm{km\,s^{-1}}$. The laboratory velocity is
$v_{\mathrm{lab}} = v_{\odot} + v_{\oplus} + v_R$, with
$v_{\odot} \simeq 230\,\mathrm{km\,s^{-1}}$ the solar velocity relative
to the Galactic frame, $v_{\oplus} \simeq 29.8\,\mathrm{km\,s^{-1}}$
Earth's orbital speed, and $v_R = \omega_{\oplus} R_{\oplus}
\cos\phi_\mathrm{det}$ Earth's rotation at detector latitude
$\phi_\mathrm{det}$, where $\omega_{\oplus} = 2\pi/\mathrm{day}$ and
$R_{\oplus} \simeq 6371$~km~\cite{PhysRevD.96.123008}.

Filter-induced correlations between neighbouring GS bins were corrected
to first order by dividing by the measured standard deviation, evaluated
over a line-free window. The GS was normalized by a robust
median-absolute-deviation estimate of the noise scale over the
instrumentally unmasked band, so that the detection threshold
$\alpha=3$ corresponds to exactly $3\sigma$. Across twenty equal
sub-intervals of the band the median of the normalized excess varies
with an rms of $0.09\,\sigma$; this residual baseline structure is
carried as a systematic uncertainty on the reported limit.


Narrow spectral features that survive the upstream screening can become
significant only at the GS stage, where persistent lines accumulate
through stacking, rebinning and line-shape weighting. Fixed instrumental
masks and bins associated with narrow IF/RF interferences are excluded
before candidate selection. Positive excesses above $\alpha=3$ are
grouped into contiguous clusters; those with widths compatible with the
expected dark-photon linewidth ($3\leq N_{\rm bins}\leq7$) are
provisionally retained, while wider clusters are vetoed as broad noise
or systematics.

Negative excursions are vetoed symmetrically with the positive side.
Baseline removal turns any narrow feature into a residual with negative
sidelobes, so a strong spurious line imprints companion excursions of
opposite sign. These sit more than twice as close to the positive
excesses as random placement would allow, confirming that they trace the
same instrumental contamination that generates them. Both categories are
excluded using adaptive veto windows whose extent is set by the local
profile of the excess.

\begin{figure}[h]
    \centering    \includegraphics[width=.48\textwidth]{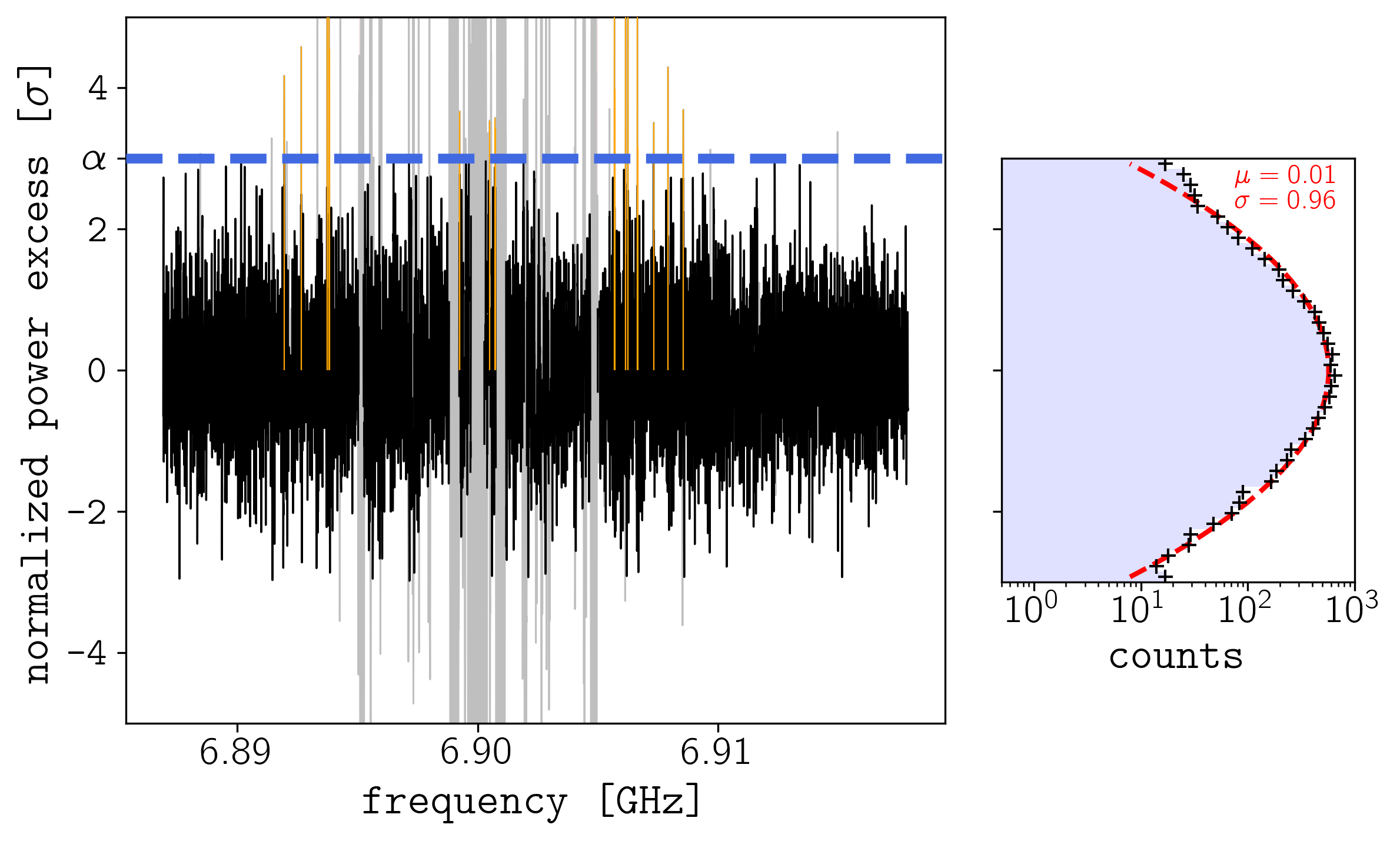}
    \caption{Left: normalized, correlation-corrected grand spectrum. Black, retained bins;
grey, masked bins; orange, positive clusters flagged for rescan
(Table~\ref{TableII}). The blue dashed line is the detection threshold
$\alpha=3$. Right: histogram of the retained normalized power excesses against
a unit normal reference (red dashed). The spectrum is normalized by a median estimate of the noise scale over the instrumentally unmasked band, so
that $\alpha=3$ corresponds to $3\sigma$; the retained bins give
$\hat\sigma=0.96$, the narrowing expected from vetoing positive and negative
excursions symmetrically.}
\label{fig_1}
\end{figure}

The end-to-end response was calibrated by injecting the laboratory-frame
dark-photon template into the recorded spectra \emph{before} the
baseline is estimated, so that the response of the baseline to the
signal itself is included. Injections at five frequencies spanning the
band and at three couplings give a pipeline detection efficiency
$\eta_{\rm e2e}\simeq0.60$, with no measurable dispersion in either
variable. This factorizes as $\eta_{\rm e2e} = \eta_{\rm match} \times
\eta_{\rm SG} \simeq 0.76 \times 0.79$, where $\eta_{\rm SG}$ is the
attenuation of the SG operator and $\eta_{\rm match}$ the residual
mismatch between the search template and the filtered signal shape. The
quoted efficiency refers exclusively to the analysis chain; instrumental
losses are accounted for separately.

Surviving clusters were subjected to two independent tests. The first is
a line-shape test on the coadded, unrebinned spectrum: we fit a Gaussian
of free width, propagated through the same baseline-removal operator as
the data, and compare it with the width of the dark-photon template
propagated identically. The test is applied only where the local excess
exceeds five times the median-absolute-deviation noise level of the
coadded spectrum, and is otherwise declared inconclusive rather than
used to veto. Second, excesses implying a kinetic mixing already
excluded at 95\% confidence by existing dark-photon searches in this
mass range are flagged and removed from further consideration.

Persistent narrow lines near 6.9~GHz were also observed with the RF
input from the cryogenic front-end disconnected and the room-temperature
chain terminated with a matched load, and in control runs with the RF
center frequency shifted by $+50$~MHz, supporting their instrumental
classification. Fourteen of the forty-six width-selected clusters
satisfy the retention criteria with a local confidence level
CL$_{\rm loc}>1\%$; these are listed in Table~\ref{TableII}. Masked
regions appear as grey traces and the retained rescan candidates as
orange traces in Fig.~\ref{fig_1}.

\begin{table}[h]
  \centering
\caption{Positive-excess clusters retained for follow-up rescans, listing frequency
range and width in grand-spectrum bins. CL$_{\rm loc}$ is the local confidence
level at which a dark photon of the corresponding signal strength is excluded
at that cluster; all values fall below the 90\% adopted elsewhere, so these
frequencies are removed from the exclusion region.}
\label{TableII}
\begin{tabular}{ccc}
\hline\hline
Frequency [GHz] & $N_{\rm bins}$ & CL$_{\rm loc}$ [\%] \\
  \hline
  6.891944--6.891950 & 3 & 54.2 \\
  6.892648--6.892656 & 4 & 38.1 \\
  6.893729--6.893741 & 5 & 17.3 \\
  6.893818--6.893829 & 5 & 5.3 \\
  6.899245--6.899260 & 6 & 72.9 \\
  6.900493--6.900499 & 3 & 77.2 \\
  6.900730--6.900739 & 4 & 75.9 \\
  6.905700--6.905711 & 5 & 16.5 \\
  6.906163--6.906175 & 5 & 7.4 \\
  6.906252--6.906263 & 5 & 5.2 \\
  6.906655--6.906667 & 5 & 3.6 \\
  6.907334--6.907339 & 3 & 77.9 \\
  6.907929--6.907937 & 4 & 49.4 \\
  6.908564--6.908572 & 4 & 72.1 \\
  \hline\hline
\end{tabular}
\end{table}

\begin{figure}[H]
    \centering    \includegraphics[width=0.48\textwidth]{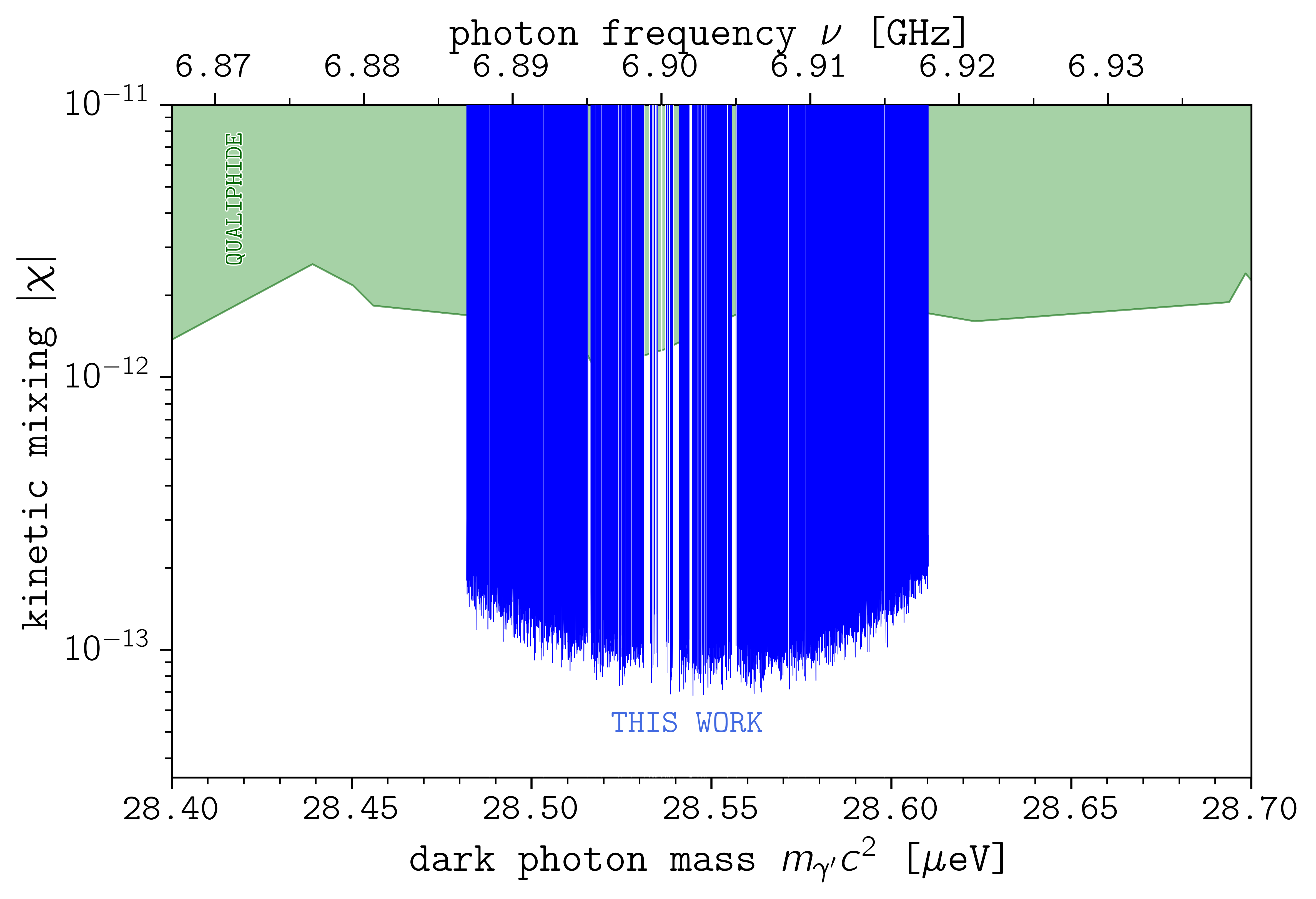}
    \caption{New exclusion region (blue) at 90\% CL. Narrow gaps indicate masked bins and vetoed neighborhoods around above-threshold excursions. Green: existing laboratory
constraints overlapping this mass range \cite{Ramanathan:2022egk}, quoted at the confidence
levels of the original publication. The figure was produced using \cite{ciaran_o_hare_2020_3932430}.}
    \label{fig_2}
\end{figure}

\noindent\textbf{Results.} 
Figure~\ref{fig_2} shows the main outcome. We analyzed 12960 files of 10.0~s
(36.0~h in total) spanning 6.886909--6.917922~GHz
(28.4820--28.6102~$\upmu$eV), reaching a sensitivity
$\chi\lesssim 6.7967 \times10^{-14}$ at 6.902112~GHz (28.5448~$\upmu$eV), sustained
over a full $K_g$-bin resolution element and
$\kappa^2=1/3$. Dark photons are excluded at 90\%
CL throughout the unmasked region, 9555 of 10841 GS bins
(88.1\%). To our knowledge, this result establishes the strongest laboratory-based exclusion limit in this frequency range. The candidates listed in Table~\ref{TableII} will be re-examined in a
future run with upgraded instrumentation.

\textit{Acknowledgements.} The project that gave rise to these results received the support of a fellowship from “la Caixa” Foundation (ID 100010434). The fellowship code is LCF/BQ/PI24/12040023”. J.DM. acknowledges support from the Spanish Ministry of Science, Innovation and Universities and the Agency (EUR2024-153552 financed by MICIU/AEI/10.13039/501100011033). We gratefully acknowledge financial support from the Severo Ochoa Program for Technological Projects and Major Surveys 2020-2023 under Grant No. CEX2019-000920-S; Recovery, Transformation and Resiliency Plan of Spanish Government under Grant No. C17.I02.CIENCIA.P5; Operational Program of the European Regional Development Fund (ERDF) under Grant No. EQC2019-006548-P; IAC Plan de Actuación 2022. This work was supported by the RIKEN's program for Special Postdoctoral Researchers (SPDR). This work is part of grant CEX2025-001609-S, awarded to the Instituto de Astrofísica de Canarias under the Severo Ochoa Centre of Excellence program and funded by MICIU/AEI/10.13039/501100011033. M.M. acknowledges funding by the Deutsche Forschungsgemeinschaft (DFG, German Research Foundation) under Germany’s Excellence Strategy---EXC 2121 "Quantum Universe"---390833306. M.E.T. was funded by the Australian Research Council Centre of Excellence for Dark Matter Particle Physics (CE200100008).

\textit{Data availability.} The source data and code used to generate the plots in this paper are available from the corresponding author upon reasonable request.


\bibliography{apssamp}

@ARTICLE{1946RScI...17..268D,
       author = {{Dicke}, R.~H.},
        title = "{The Measurement of Thermal Radiation at Microwave Frequencies}",
      journal = {Review of Scientific Instruments},
         year = 1946,
        month = jul,
       volume = {17},
       number = {7},
        pages = {268-275},
          doi = {10.1063/1.1770483},
       adsurl = {https://ui.adsabs.harvard.edu/abs/1946RScI...17..268D}
}

@ARTICLE{1970ApJ...159..379R,
       author = {{Rubin}, Vera C. and {Ford}, W. Kent, Jr.},
        title = "{Rotation of the Andromeda Nebula from a Spectroscopic Survey of Emission Regions}",
      journal = {Astrophysical Journal},
         year = 1970,
        month = feb,
       volume = {159},
        pages = {379},
          doi = {10.1086/150317},
       adsurl = {https://ui.adsabs.harvard.edu/abs/1970ApJ...159..379R}
}

@ARTICLE{1983PhRvL..51.1415S,
       author = {{Sikivie}, P.},
        title = "{Experimental Tests of the ``Invisible'' Axion}",
      journal = {Phys. Rev. Lett.},
         year = 1983,
        month = oct,
       volume = {51},
       number = {16},
        pages = {1415-1417},
          doi = {10.1103/PhysRevLett.51.1415},
       adsurl = {https://ui.adsabs.harvard.edu/abs/1983PhRvL..51.1415S}
}

@article{
doi:10.1126/sciadv.abq3765,
author = {Aaron Quiskamp  and Ben T. McAllister  and Paul Altin  and Eugene N. Ivanov  and Maxim Goryachev  and Michael E. Tobar },
title = {Direct search for dark matter axions excluding ALP cogenesis in the 63 to 67 $\mu$eV range with the ORGAN experiment},
journal = {Science Advances},
volume = {8},
number = {27},
pages = {eabq3765},
year = {2022},
doi = {10.1126/sciadv.abq3765},
URL = {https://www.science.org/doi/abs/10.1126/sciadv.abq3765},
eprint = {https://www.science.org/doi/pdf/10.1126/sciadv.abq3765}}

@article{Okun:1982xi,
    author = "Okun, L. B.",
    title = "{Limits of electrodynamics: paraphotons?}",
    reportNumber = "ITEP-48-1982",
    journal = "Sov. Phys. JETP",
    volume = "56",
    pages = "502",
    year = "1982"
}

@article{Caputo:2021eaa,
    author = "Caputo, Andrea and Millar, Alexander J. and O'Hare, Ciaran A. J. and Vitagliano, Edoardo",
    title = "{Dark photon limits: A handbook}",
    eprint = "2105.04565",
    archivePrefix = "arXiv",
    primaryClass = "hep-ph",
    reportNumber = "NORDITA-2021-036",
    doi = "10.1103/PhysRevD.104.095029",
    journal = "Phys. Rev. D",
    volume = "104",
    number = "9",
    pages = "095029",
    year = "2021"
}

@article{PhysRevD.96.123008,
  title = "{HAYSTAC axion search analysis procedure}",
  author = {Brubaker, B. M. and Zhong, L. and Lamoreaux, S. K. and Lehnert, K. W. and van Bibber, K. A.},
  journal = {Phys. Rev. D},
  volume = {96},
  issue = {12},
  pages = {123008},
  numpages = {34},
  year = {2017},
  month = {Dec},
  publisher = {American Physical Society},
  doi = {10.1103/PhysRevD.96.123008},
  url = {https://link.aps.org/doi/10.1103/PhysRevD.96.123008}
}

@misc{ciaran_o_hare_2020_3932430,
  author       = {Ciaran O'Hare},
  title        = {cajohare/AxionLimits: AxionLimits},
  month        = jul,
  year         = 2020,
  publisher    = {Zenodo},
  version      = {v1.0},
  doi          = {10.5281/zenodo.3932430},
  url          = {https://doi.org/10.5281/zenodo.3932430}
}

@ARTICLE{1899ApJ.....9...87P,
       author = {{Perot}, A. and {Fabry}, Charles},
        title = "{On the Application of Interference Phenomena to the Solution of Various Problems of Spectroscopy and Metrology}",
      journal = {Astrophysical Journal},
         year = 1899,
        month = feb,
       volume = {9},
        pages = {87},
          doi = {10.1086/140557},
       adsurl = {https://ui.adsabs.harvard.edu/abs/1899ApJ.....9...87P}
}

@article{Ramanathan:2022egk,
    author = "Ramanathan, Karthik and Klimovich, Nikita and Basu Thakur, Ritoban and Eom, Byeong Ho and LeDuc, Henry G. and Shu, Shibo and Beyer, Andrew D. and Day, Peter K.",
    title = "{Wideband Direct Detection Constraints on Hidden Photon Dark Matter with the QUALIPHIDE Experiment}",
    eprint = "2209.03419",
    archivePrefix = "arXiv",
    primaryClass = "astro-ph.CO",
    doi = "10.1103/PhysRevLett.130.231001",
    journal = "Phys. Rev. Lett.",
    volume = "130",
    number = "23",
    pages = "231001",
    year = "2023"
}

@article{Staudt:2024tdq,
    author = "Staudt, Patrick G. and Bullock, James S. and Boylan-Kolchin, Michael and Wetzel, Andrew and Ou, Xiaowei",
    title = "{Sliding into DM: Determining the local dark matter density and speed distribution using only the local circular speed of the Galaxy}",
    eprint = "2403.04122",
    archivePrefix = "arXiv",
    primaryClass = "astro-ph.GA",
    month = "3",
    year = "2024",
    journal = {arXiv:2403.04122}
}

@article{PhysRevD.110.072013,
  title = {Echo-free quality factor of a multilayer axion haloscope},
  author = {Hern\'andez-Cabrera, Juan F. and De Miguel, Javier and Hern\'andez-Su\'arez, E. and Joven, Enrique and Otani, Chiko and Rubi\~no-Mart\'{\i}n, J. Alberto and Zioutas, Konstantin},
  collaboration = {DALI Collaboration},
  journal = {Phys. Rev. D},
  volume = {110},
  issue = {7},
  pages = {072013},
  numpages = {11},
  year = {2024},
  month = {Oct},
  publisher = {American Physical Society},
  doi = {10.1103/PhysRevD.110.072013},
  url = {https://link.aps.org/doi/10.1103/PhysRevD.110.072013}
}

@article{DeMiguel2021,
	doi = {10.1088/1475-7516/2021/04/075},
	url = {https://doi.org/10.1088/1475-7516/2021/04/075},
	year = 2021,
	month = {apr},
	publisher = {{IOP} Publishing},
	volume = {2021},
	number = {04},
	pages = {075},
	author = {{De Miguel}, Javier},
	title = {A dark matter telescope probing 
		the 6 to 60 {GHz} band},
	journal = {Journal of Cosmology and Astroparticle Physics}
}

@article{DeMiguel:2023nmz,
    author = "De Miguel, Javier and Hern{\'a}ndez-Cabrera, Juan F. and Hern{\'a}ndez-Su{\'a}rez, Elvio and Joven-{\'A}lvarez, Enrique and Otani, Chiko and Rubi{\~n}o-Mart{\'\i}n, J. Alberto",
    collaboration = "DALI",
    title = "{Discovery prospects with the Dark-photons {\&} Axion-like particles Interferometer}",
    eprint = "2303.03997",
    archivePrefix = "arXiv",
    primaryClass = "hep-ph",
    doi = "10.1103/PhysRevD.109.062002",
    journal = "Phys. Rev. D",
    volume = "109",
    number = "6",
    pages = "062002",
    year = "2024"
}

@article{Cabrera2023qkt,
    author = "Hern\'andez-Cabrera, Juan F. and De Miguel, Javier and \'Alvarez, Enrique Joven and Hern\'andez-Su\'arez, E. and Rubi\~no-Mart\'in, J. Alberto and Otani, Chiko",
    title = "{A Forecast of the Sensitivity of the DALI Experiment to Galactic Axion Dark Matter}",
    eprint = "2310.20437",
    archivePrefix = "arXiv",
    primaryClass = "hep-ph",
    doi = "10.3390/sym16020163",
    journal = "Symmetry",
    volume = "16",
    number = "2",
    pages = "163",
    year = "2024"
}

@article{2024JInst19P1022H,
       author = {{Hern{\'a}ndez-Cabrera}, Juan F. and {De Miguel}, Javier and {Hern{\'a}ndez-Su{\'a}rez}, E. and {Joven-{\'A}lvarez}, Enrique and {Lorenzo-Hern{\'a}ndez}, H. and {Otani}, Chiko and {Rapado-Tamarit}, Miguel A. and {Rubi\~no-Mart\'in}, J. Alberto},
        title = "{Experimental measurement of the quality factor of a Fabry-P{\'e}rot open-cavity axion haloscope}",
      journal = {Journal of Instrumentation},
         year = 2024,
        month = jan,
       volume = {19},
       number = {1},
          eid = {P01022},
        pages = {P01022},
          doi = {10.1088/1748-0221/19/01/P01022},
archivePrefix = {arXiv},
       eprint = {2310.16013},
 primaryClass = {hep-ph},
       adsurl = {https://ui.adsabs.harvard.edu/abs/2024JInst..19P1022H}
}

@article{DeMiguel:2024cwb,
    author = "De Miguel, Javier and Kryemadhi, Abaz and Zioutas, Konstantin",
    collaboration = "DALI",
    title = "{DALI sensitivity to streaming axion dark matter}",
    eprint = "2404.13970",
    archivePrefix = "arXiv",
    primaryClass = "hep-ph",
    doi = "10.1103/PhysRevD.111.023016",
    journal = "Phys. Rev. D",
    volume = "111",
    number = "2",
    pages = "023016",
    year = "2025"
}

@ARTICLE{1386650,
  author={Frigo, M. and Johnson, S.G.},
  journal={Proceedings of the IEEE}, 
  title={The Design and Implementation of FFTW3}, 
  year={2005},
  volume={93},
  number={2},
  pages={216-231},
  doi={10.1109/JPROC.2004.840301}}

@article{DeMiguel:2026mvi,
    author = "De Miguel, Javier and others",
    title = "{First Limits on Axion Dark Matter from a DALI Prototype}",
    eprint = "2603.21951",
    archivePrefix = "arXiv",
    primaryClass = "hep-ex",
    month = "3",
    year = "2026",
    journal = {arXiv:2603.21951}
}






\end{document}